\documentclass[conference]{IEEEtran}
\IEEEoverridecommandlockouts

\usepackage{cite}
\usepackage{amsmath,amssymb,amsfonts}
\usepackage{algorithmic}
\usepackage{graphicx}
\usepackage{textcomp}
\usepackage{xcolor}

\usepackage[caption=false,font=footnotesize]{subfig}

\makeatletter

\renewcommand{\p@subfigure}{\thefigure}
\makeatother

\usepackage{array}
\usepackage{balance}
\usepackage[ruled,linesnumbered]{algorithm2e}  
\usepackage{booktabs}   
\usepackage{multirow}   
\usepackage{siunitx}   
\usepackage{textcomp}
\usepackage{cleveref}
\usepackage{tikz}
\usepackage{xcolor}
\usepackage[most]{tcolorbox}
\def\BibTeX{{\rm B\kern-.05em{\sc i\kern-.025em b}\kern-.08em
    T\kern-.1667em\lower.7ex\hbox{E}\kern-.125emX}}
\begin{document}
\title{\textsc{Uni-Fi}: Integrated Multi-Task Wi-Fi Sensing\vspace{-2mm}}
\author{
Mengning Li and Wenye Wang%
\thanks{This work was supported in part by Institute for Connected Sensor-Systems - NC State University.}
\\
\IEEEauthorblockA{NC State University}
\IEEEauthorblockA{mli55@ncsu.edu, wwang@ncsu.edu}
}

\maketitle

\begin{abstract}
Wi-Fi sensing technology enables non-intrusive, continuous monitoring of user locations and activities, which supports diverse smart home applications.
Since different sensing tasks exhibit contextual relationships, their integration can enhance individual module performance.
However, integrating sensing tasks across different studies faces challenges due to the absence of: 1) a unified architecture that captures the fundamental nature shared across diverse sensing tasks, and 2) an extensible pipeline that accommodates future sensing methodologies.
This paper presents \textsc{Uni-Fi}, an extensible framework for multi-task Wi-Fi sensing integration.
This paper makes the following contributions:
1) we propose a unified theoretical framework that reveals fundamental differences between single-task and multi-task sensing;
2) we develop a scalable sensing pipeline that automatically generates a multi-task sensing solver, enabling seamless integration of multiple sensing models.
Experimental results show that \textsc{Uni-Fi} achieves robust performance across tasks, with a median localization error of approximately $0.54\,\mathrm{m}$, $98.34\%$ accuracy for activity classification, and $98.57\%$ accuracy for presence detection.
\end{abstract}


\section{Introduction}
\subsection{Background and Motivation}
Wi-Fi sensing technology enables non-intrusive, continuous monitoring of user location and status, thereby facilitating diverse sensing tasks for smart homes.
Existing research has demonstrated its potential in various applications including activity recognition\cite{liu2023towards,li2024hybrid,liu2024unifi,gao2022towards,shi2024investigation}, user tracking \cite{qian2017widar,qian2018widar2,tong2024nne,tong2021mapfi,xu2023hypertracking,wu2021witraj}, presence detection\cite{jayaweera2024robust,shen2024time,wei2025survey,du2022crcloc,ge2023crosstrack,yu2025lowdetrack}, and respiration monitoring\cite{chang2024robust,zeng2019farsense,zeng2020multisense,wilson2010seethrough}.
By persistently observing these user states, we can effectively profile behavioral patterns, ultimately advancing the intelligence of next-generation home automation systems.

Integrating multiple Wi-Fi sensing tasks can significantly enhance the performance of individual sensing modules\cite{tong2025stagr}.
Two typical examples are as follows:
1) While Wi-Fi sensing alone struggles with fine-grained activity characterization, incorporating location context (\textit{e.g.}, a user in a specific rooms) enables more precise behavior inference;
2) Standalone presence detection suffers from false alarms, but analyzing spatial relationships between current users' positions and room entrances can substantially improve detection reliability.

\begin{figure}[t]
    \centering
    \subfloat[Analyzing spatial relationships between current users’ position and room entrances can substantially improve detection reliability.]{%
        \includegraphics[width=0.99\linewidth]{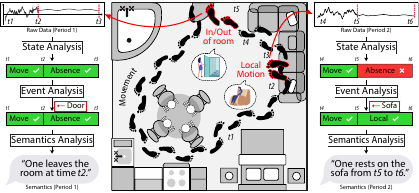}%
        \label{fig:moti1}
    }
    \vspace{-1mm}
    
    \subfloat[Single-task sensing vs. multiple-task sensing.]{%
        \includegraphics[width=0.99\linewidth]{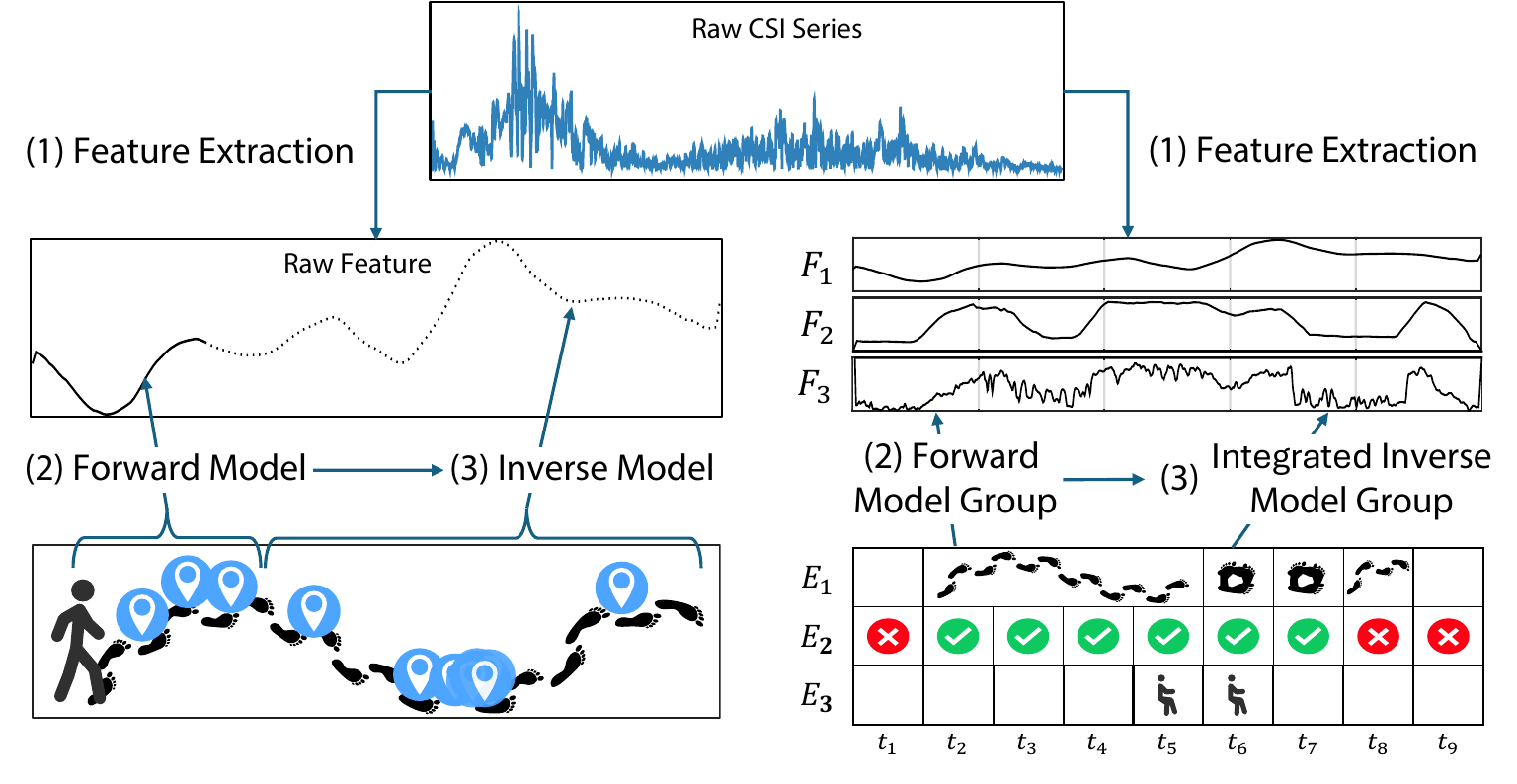}%
        \label{fig:moti2}
    }
    \caption{Rationale behind Uni-Fi for multi-task sensing.}
    \vspace{-2mm}
    \label{fig:motivation}\vspace{-5mm}
\end{figure}

Current research has made substantial efforts in improving individual sensing tasks, yet integrating sensing tasks proposed across different studies remains non-trivial.
As illustrated in Fig.~\ref{fig:motivation}, the sensing paradigm operates through three fundamental stages:
1) We first specify the sensing task, then derive discriminative features through theoretical modeling and experimental validation;
2) We establish the mapping relationship (\textit{Forward Models}) that characterizes how the target task influences these observable features;
3) We transform the mapping relationship into executable algorithms (\textit{Inverse Models}) that enable task inference from observed feature patterns, thereby deploying the sensing system.
However, the introduction of novel sensing tasks or the discovery of new sensing features \cite{xu2023hypertracking,wang2022placement,li2024hybrid,ma2018signfi,xie2019mdtrack} necessitates the redesign of inverse models, causing a fundamental scalability challenge in developing multi-task integrated systems.

\begin{figure*}[t]
    \centering
    \subfloat[Absence condition.]{\includegraphics[width=0.24\linewidth]{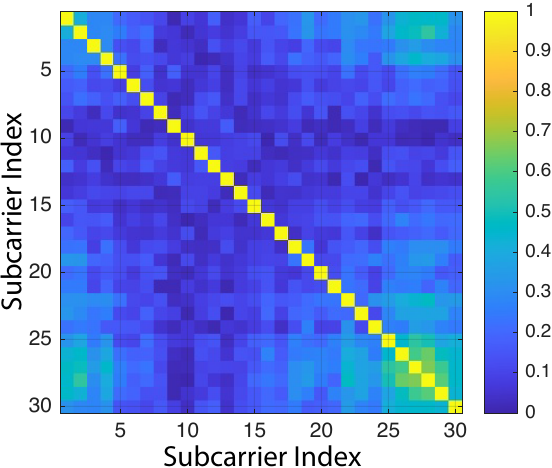}\label{fig:prelim1a}}
    \subfloat[Static breathing.]{\includegraphics[width=0.24\linewidth]{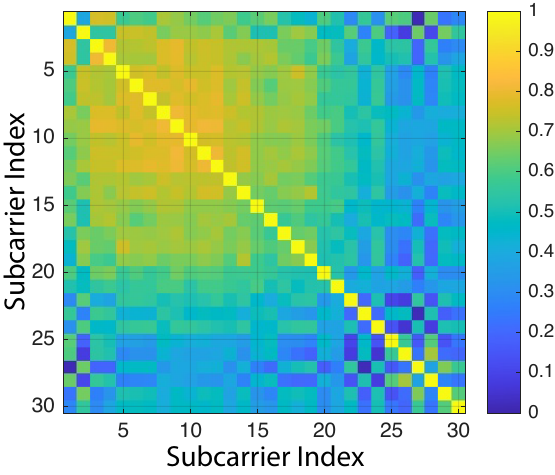}\label{fig:prelim1b}}
    \subfloat[PLCR comparison.]
    {\includegraphics[width=0.24\linewidth]{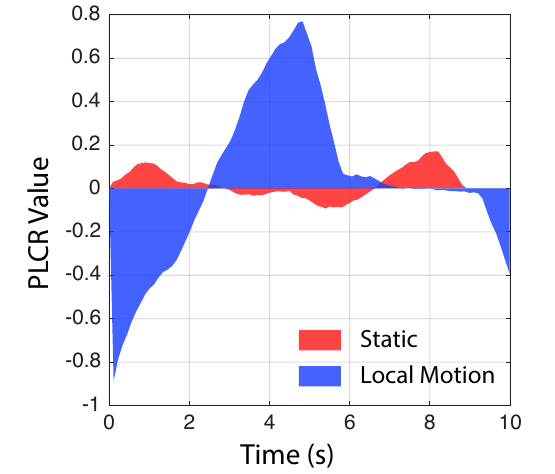}\label{fig:prelim2}}
    \subfloat[DSER across behaviors.]{\includegraphics[width=0.24\linewidth]{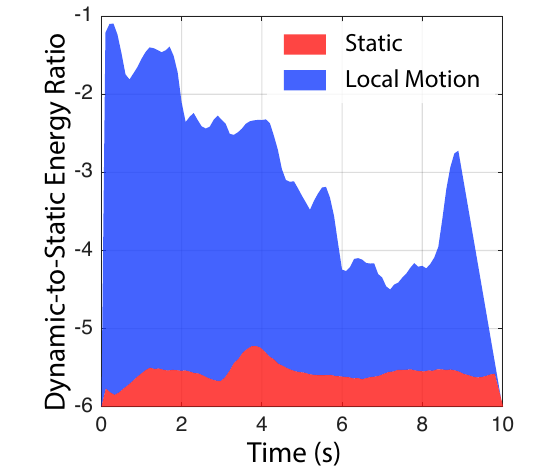}\label{fig:prelim3}}
    \caption{Comparison of features across behaviors between absence and static breathing.}
\vspace{-5mm}
\end{figure*}

\subsection{The Basic Idea of \textsc{Uni-Fi}}
This paper presents \textsc{Uni-Fi}, an extensible multi-task integrated framework for Wi-Fi sensing.
\textsc{Uni-Fi} implements an automated pipeline that aggregates forward models from diverse sensing tasks into executable aggregated inverse models.
The workflow is as follows:
first, we define the event set and feature set to establish the boundaries for sensing tasks;
second, through synthetic event sequences, we model user activities, which are then transformed into feature sequences via forward models to construct the dataset;
third, we utilize the data-driven approach to learn the mapping between feature sequences and event sequences, thereby building executable inverse models.
Designing such an integrated multi-task Wi-Fi sensing system presents non-trivial challenges, including:




\textit{Challenge 1: Unified Formalization of Fundamental Principles in Wi-Fi Sensing.}
A fundamental challenge in developing \textsc{Uni-Fi} stems from the need to establish a unified theoretical framework that can systematically integrate diverse sensing tasks from existing literature.
The core difficulty lies in identifying whether current wireless sensing methodologies adhere to a consistent technical paradigm, and subsequently characterizing its essential components.
This undertaking is particularly non-trivial due to both the extensive corpus of published research and the inherent heterogeneity in sensing objectives and methodological approaches across studies.

\textit{Our Insight:}
We reveal that the essence of single-task sensing lies in establishing the mapping relationship between features and the task as forward models, and then transforming such forward models into executable functions called inverse models.
In multi-task sensing, paired feature and task sets are interconnected through a network composed of multiple forward models (forward model network).
The core objective is to derive the aggregated inverse model that performs set-to-set transformation between features and events.


\textit{Challenge 2: Scalable Pipeline for Integrating Additional Sensing Tasks.}
A multi-task sensing system must demonstrate strong scalability to effectively incorporate new valuable research findings.
For instance, when novel Wi-Fi sensing tasks are discovered or more effective sensing features are identified, the system should be capable of rapidly integrating these new developments.
However, designing a framework with sufficient extensibility to accommodate unknown future research directions presents significant challenges.

\textit{Our Insight:}
In our multi-task sensing framework, integrating new sensing tasks can be viewed as modifying the boundaries of either the task set or feature set, along with changes to the forward model network. 
Specifically, when new tasks or features are added, the feature set and task set will be connected through an updated model network that incorporates both the original forward models and newly proposed forward models from recent studies.
We present a unified pipeline capable of data-driven neural network representation for the forward model network, enabling new research to be integrated within the existing framework while enhancing scalability.

\subsection{Contributions}
The key contributions are as follows:
\begin{itemize}
  \item \textit{Unified Theoretical Framework:} We establish the first unified theoretical framework for Wi-Fi sensing by formalizing single-task sensing as feature-to-task mappings (forward models) and their executable implementations (inverse models). This reveals multi-task sensing requires interconnected feature/event sets through forward model networks, solved via aggregated inverse models.
  \item \textit{Scalable Sensing Pipeline:} We design a scalable pipeline where new tasks/features extend predefined sets and forward model networks. Our data-driven approach enables neural networks to dynamically represent updated model networks, allowing seamless integration of novel research while maintaining framework consistency.
  \item \textit{Extensive Real-world Evaluations:} We conduct experiments and show that our system maintains stable performance across diverse environments and tasks, achieving a localization error of approximately $0.54m$ and $98.34\%$ accuracy for posture classification.
\end{itemize}

\begin{table*}[t]
\centering
\caption{Definitions of core concepts in \textsc{Uni-Fi}}
\label{tab:concepts}
\begin{tabular}{>{\centering\arraybackslash}m{3.2cm} >{\centering\arraybackslash}m{1.4cm} >{\centering\arraybackslash}m{12cm}}
\toprule
\textbf{Concept} & \textbf{Category} & \textbf{Definition} \\ \hline
\textit{Task} & Single-Task & A sensing objective, such as \emph{tracking}, \emph{respiration monitoring}, or \emph{state recognition}. \\ 
\textit{Event} & Single-Task & A primitive human state observed by the radio channel, \textit{e.g.}, \emph{nobody}, \emph{stillness}, \emph{local motion}, or \emph{walking}. \\ 
\textit{Feature} & Single-Task & A descriptor extracted directly from raw data, \textit{e.g.}, \emph{subcarrier correlation}, \emph{DSER}, and \emph{PLCR}. \\ 
\textit{Forward Model} & Single-Task & Analytical equations that map an event to its expected feature signature (existing theoretical models). \\ 
\textit{Inverse Model} & Single-Task & The proposed model that maps features back to events and, by composition, to task-level labels. \\ 
\textit{Forward Model Network} & Multi-Task & An interconnected network of forward models that transform event sets into feature sets. \\ 
\textit{Hybrid Inverse Model} & Multi-Task & The target is to derive a hybrid inverse model transforming feature sets to corresponding event sets. \\ \bottomrule
\end{tabular}\vspace{-3mm}
\end{table*}

The remainder of the paper is organized as follows. Section~\ref{sec:prelim} introduces the theoretical foundations of Wi-Fi sensing. Section~\ref{sec:insights} presents key observations that motivate our design. Section~\ref{sec:ubifidesign} describes the unified system architecture. Section~\ref{sec:experiment} reports experimental results. Section~\ref{sec:related} reviews related work, and Section~\ref{sec:conclude} concludes the paper.

\section{Preliminaries}\label{sec:prelim}
\subsection{Channel State Information}

Channel State Information (CSI) encapsulates the wireless channel characteristics between a transmitter and receiver. The static component, denoted as $H_s$, reflects stable attributes such as antenna gain patterns and large-scale path loss. In contrast, the dynamic component $H_d$ captures temporal fluctuations arising from environmental dynamics such as human movement.
Mathematically, the CSI at frequency $f$ and time $t$ can be expressed as
\begin{equation}\label{eq:csisignal}
H(f,t) = H_s(f,t) + H_d(f,t) = H_s(f,t) + Ae^{-j2\pi\frac{d(t)}{\lambda}},
\end{equation}
where $A$ accounts for complex attenuation, $e^{-j2\pi\cdot d(t)/\lambda}$ represents the phase shift due to the path length $d(t)$, and $\lambda$ is the signal wavelength.
This formulation shows how environmental variations influence wireless propagation.
\subsection{Subcarrier Correlation}
The subcarrier correlation serves as a valuable metric for identifying the presence of humans.
Using the decomposition $H(f,t) = H_s(f,t) + H_d(f,t)$, the subcarrier correlation matrix $\mathbf{R}$ captures the coherence of the frequency domain:
\begin{equation}
\mathbf{R}(f_i,f_j) = \mathbb{E}[H(f_i,t)H^*(f_j,t)].
\end{equation}
$\mathbf{R}$ exhibits distinct characteristics depending on occupancy.
As illustrated in Fig.~\ref{fig:prelim1a} and Fig.~\ref{fig:prelim1b}, unoccupied scenarios are characterized by:  
1) uniformly low correlation maps, primarily dominated by $H_s$; and  
2) in contrast, occupied environments exhibit high-energy, structured patterns due to the contribution of $H_d$.
This empirical observation confirms subcarrier correlation as an indicator of human presence.

\subsection{Dynamic-to-Static Energy Ratio}
The Dynamic-to-Static Energy Ratio (DSER) quantifies the strength of motion-induced dynamics in Wi-Fi signals and serves as an effective indicator of motion detection.
It distinguishes between static and active states by isolating the dynamic component from other sources of signal energy.
DSER is defined as
\begin{equation}
\text{DSER}(f,t) = \log \frac{|H_d(f,t)|^2}{|H_s(f,t)|^2}.
\end{equation}
Fig.~\ref{fig:prelim2} compares the temporal DSER profiles in static and dynamic settings.
In the absence of human activity, the DSER remains low and stable, reflecting minimal signal fluctuation.
In contrast, the presence of human motion (e.g., walking or gesturing) introduces a distinct elevation in SSNR~\cite{wang2023dfsense} magnitude and temporal variability.
This disparity confirms DSER as a reliable metric for distinguishing active occupancy from background states.

\subsection{Path-Length Change Rate}
The Path-Length Change Rate (PLCR) enables effective discrimination between stationary user actions and user walking.
The Doppler frequency introduced by motion along a propagation path is given by
\begin{equation}
f_D(t) = -\frac{\text{d}}{\text{d} t} \frac{d(t)}{\lambda}.
\end{equation}
PLCR corresponds to the derivative term $\frac{\text{d}}{\text{d}t} d(t)$, reflecting how rapidly the signal path changes due to motion.
Fig.~\ref{fig:prelim3} illustrates PLCR sequences under two behavioral states: local motion (\textit{e.g.}, hand gesture), and movement.
Local motion produces low-amplitude, short-duration PLCR spikes, while continuous locomotion results in sustained, high-magnitude changes. These patterns enable clear differentiation of motion intensity and temporal structure.


\section{Basic Idea}\label{sec:insights}
\begin{figure*}[t]
    \centering
    \includegraphics[width=.97\linewidth]{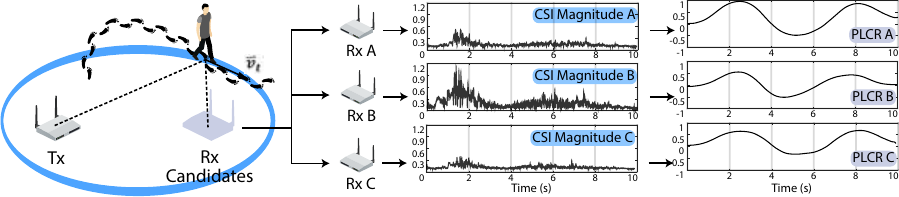}
    \caption{Answer to RQ1: The feature-level information demonstrates enhanced stability and scalability.}\vspace{-3mm}
    \label{fig:observe1}
\end{figure*}

\begin{figure*}[t]
    \centering
    \includegraphics[width=1\linewidth]{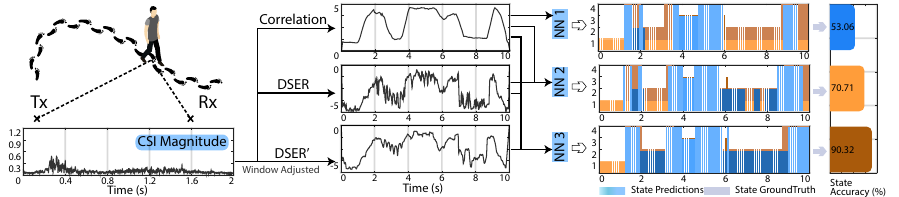}
    \caption{Answer to RQ2: Different features or observations with varying window lengths exhibit complementary advantages.}
    \label{fig:observe2}\vspace{-5mm}
\end{figure*}

Table \ref{tab:concepts} provides formal definitions for the core concepts underpinning the \textsc{Uni-Fi} framework. Building upon these, this section details three key observations that inform our system design and address three primary research questions (RQs).


\subsection{\textbf{RQ1:} Why do we choose feature-level information?}\label{sec:insight1}


Feature representations play a central role in enabling reliable and generalizable Wi-Fi sensing.
%
%
While raw CSI contains richer information, the origin of this information remains unclear—it may stem from device heterogeneity or unknown noise.
Although training neural networks with such data in an overfitting manner may yield favorable results, the described approach lacks scalability.
Therefore, it is essential to leverage wireless signal knowledge to extract meaningful features that advance sensing capabilities.



To validate our hypothesis, we conduct the experiment illustrated in Fig.~\ref{fig:observe1}.
All receivers are placed at the same physical location, and the transmitter remains fixed throughout the experiment to eliminate spatial variation.
\begin{itemize}
    \item Raw Information: Despite this controlled setup, the raw CSI recorded by the three receivers shows substantial differences, reflecting the impact of device-specific hardware characteristics.
    The average cross-device similarity is just 43.7\% when using raw CSI directly.
    \item Feature-Level Information: When we extract PLCR features from the raw CSI readings, the resulting feature representations exhibit strong alignment across all devices. The average cross-device similarity based on PLCR features reaches 78.1\%.
\end{itemize}

\textbf{Answer to RQ1:} Knowledge-driven feature extraction can derive valuable information from raw data. The feature-level information demonstrates enhanced stability and scalability.


\subsection{\textbf{RQ2:} Why should we construct a feature set?}\label{sec:insight1}
Individual features alone cannot adequately address all sensing tasks.
It is essential to construct a feature set including different types of features so that we can obtain a better performance.
To validate our hypothesis, we conduct the experiments to evaluate how feature types and window lengths affect classification performance across different behavioral events including \emph{absence}, \emph{stillness}, \emph{local motion}, or \emph{movement}.

As illustrated in Fig.~\ref{fig:observe2}, we first generate synthetic user event sequences through simulation, then map these events to observable features based on empirical experimental values to construct a simulated dataset.
Subsequently, three neural networks are trained with distinct input configurations:
1) Subcarrier correlation only;
2) Subcarrier correlation + DSER;
3) Subcarrier correlation + DSER with two window lengths.
Finally, we collect real-world CSI and then extract different features to analyze sensing accuracy.
The experimental results demonstrate the state recognition accuracy of $53.06\%$, $70.71\%$, and $90.32\%$, respectively.

\textbf{Answer to RQ2:} Different features or observations with varying windows exhibit complementary advantages; therefore, we construct a feature set for multi-task sensing.



\subsection{RQ3: How to design a scalable sensing architecture?}\label{sec:insight2}

Our goal is to enable easy integration of emerging features without the need to redesign or retrain entire models.
For this purpose, we propose a modular architecture that separates feature simulation from task-specific inference, allowing features to be inserted or removed as needed.

\begin{figure*}[t]
    \centering
    \subfloat[Single-task model.]{%
        \includegraphics[width=0.16\linewidth]{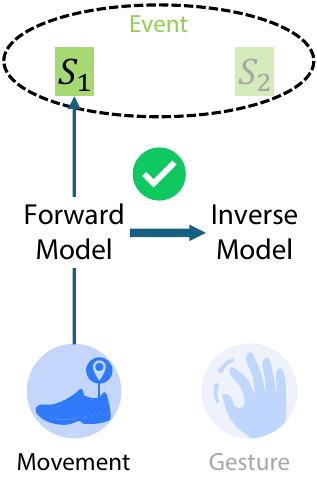}
        \label{fig:observe3:a}}
    \hfill
    \subfloat[Single-task model w.\ new applications.]{%
        \includegraphics[width=0.30\linewidth]{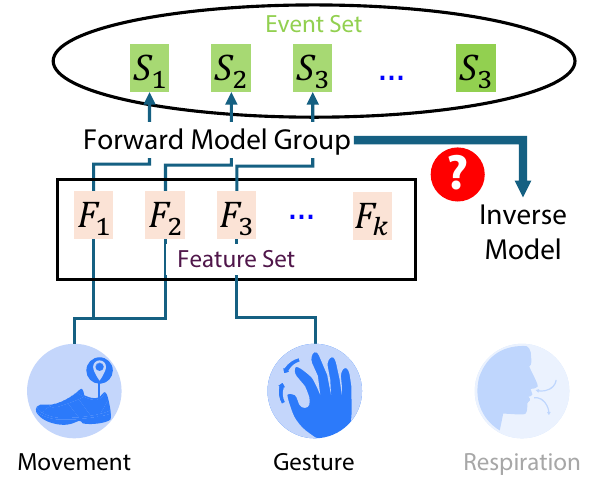}
        \label{fig:observe3:b}}
    \hfill
    \subfloat[Proposed multi-task model (\textsc{Uni-Fi}).]{%
        \includegraphics[width=0.42\linewidth]{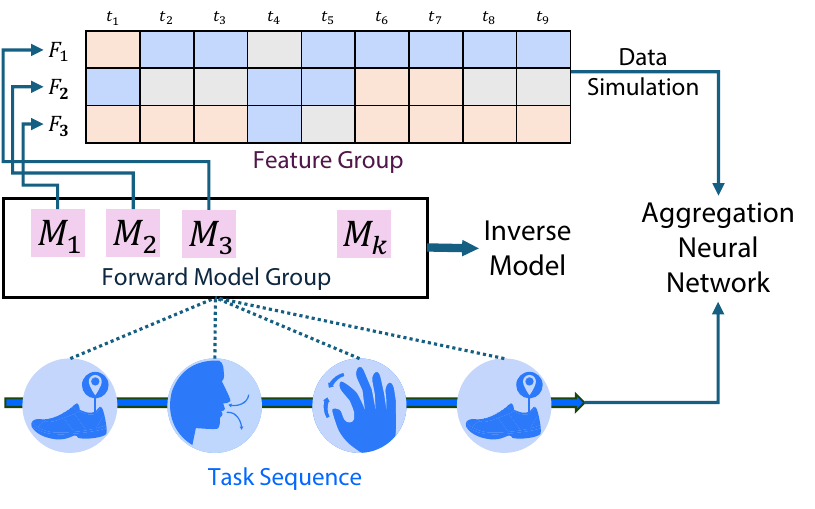}
        \label{fig:observe3:c}}
    \caption{Answer to RQ3: Decoupling feature integration from model redesign enables scalable sensing pipeline.}
    \label{fig:observe3}
    \vspace{-5mm}
\end{figure*}

The basic principle of wireless sensing is to construct the inverse model based on the forward model in Fig.~\ref{fig:observe3:a}.
However, when a new feature is discovered or a new task is integrated in Fig.~\ref{fig:observe3:b}, we need to redesign the inverse model.
This often involves designing new theoretical mechanisms to integrate the feature, such as determining its weight, interaction with other features, or compatibility with the current network. Such a tight coupling between feature selection and model architecture makes the system rigid and hard to scale.

Fig.~\ref{fig:observe3:c} illustrates the basic idea of \textsc{Uni-Fi}.
We first use a simulator to generate event sequences and then convert them into feature sequences for dataset construction.
A neural network is then trained to learn the mapping from events to features, thereby serving as the aggregated inverse model.
In this way, our pipeline enables the automatic generation of the aggregated inverse model.

\textbf{Answer to RQ3:} Decoupling feature integration from model redesign enables plug-and-play.

\section{System Design: Uni-Fi}\label{sec:ubifidesign}
\subsection{System Overview}\label{sec:system-overview}
\begin{figure*}[t]
    \centering
    \includegraphics[width=.97\linewidth]{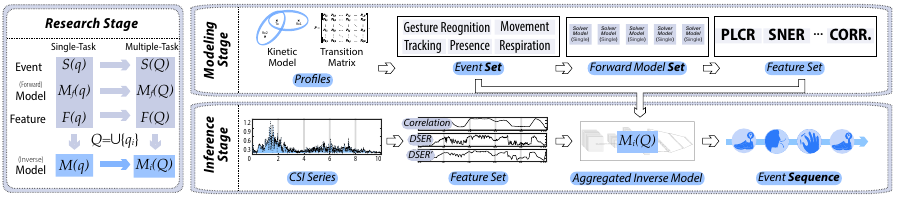}\vspace{-3mm}
    \caption{System architecture of \textsc{Uni-Fi}, consisting of research, modeling, and inference tracks.}
    \label{fig:system-overview}
\vspace{-4mm}
\end{figure*}

Fig.~\ref{fig:system-overview} presents the architecture of \textsc{Uni-Fi}.
which consists of the \textit{research} track, \textit{modeling} track, and \textit{inference} track.

\textbf{\textit{The Research Track.}} The research track primarily reveals that the essence of sensing lies in establishing a forward model from events to features, while employing an inverse model for solving.
Here, we define the boundaries of sensing, including the event and the feature set.
Furthermore, through theoretical modeling and empirical analysis, we clarify the differences between single-task sensing and multi-task sensing.

\textbf{\textit{The Modeling Track.} }
The modeling track proposes an extensible pipeline to describe how automated multi-perception task integration can be achieved.
At this stage, we introduce an event simulator to generate event sequences, which are then transformed into feature sequences through a series of forward models.
Based on the above simulated dataset, we can generate an aggregated inverse model for multi-task sensing.

\textbf{\textit{The Inference Track.} }
In the inference stage, we first extract corresponding features from the raw CSI data based on the predefined feature set.
We then apply the aggregated inverse model to transform the feature set into an event set, thereby achieving rapid inference.


\begin{table}[t]
\centering
\caption{Empirical feature ranges under different behaviors.}
\label{tab:feature-ranges-transpose}
\begin{tabular}{ccccc}
\toprule
\textbf{Feature} & \textbf{Absence} & \textbf{Stillness} & \textbf{Local Motion} & \textbf{Movement} \\
\hline
Corr. (0.5s) & 0.1 to 0.3 & 0.2 to 0.7 & 0.6 to 1.0 & 0.6 to 1.0 \\
DSER (0.5s) & -6 to -4 & -5.2 to -4 & -4 to -1 & -4 to -1 \\
PLCR (0.1s) & 0 to 0.1 & 0 to 0.1 & 0 to 0.3 & Model \\
Corr. (2s) & 0.1 to 0.3 & 0.4 to 0.7 & 0.6 to 1.0 & 0.6 to 1.0 \\
DSER (2s) & -6 to -4 & -5 to -2.5 & -4 to 0 & -4 to 0 \\
\bottomrule
\end{tabular}
\vspace{-5mm}
\end{table}

\subsection{Research Track: Unified Theoretical Framework}\label{sec:research-path}
The research track focuses on identifying which features are most useful for sensing and how they relate to specific tasks.
This part introduces three core components:
1) \textit{Event Set.} The event set $\mathcal{S}$ describes the output range of sensing results. Different sensing tasks correspond to different event sets. For instance, in gesture recognition, we define various gesture categories, while in tracking tasks, we specify the boundaries of the user's feasible area. The event set depends on the sensing task $q$;
2) \textit{Feature Set.} The feature set $\mathcal{F}$ describes how we process raw signals to accomplish the sensing task.
Even for the same sensing task, different features can be selected, and the quality of these features directly determines the accuracy of sensing;
3) \textit{Forward model.} The forward model $\mathcal{M}_f$ describes how sensing events influence features, (i.e., the mapping from events to features). 

For single-task sensing, we define the event set $\mathcal{S}(q)$ and feature set $\mathcal{F}(q)$ corresponding to task $q$.
We then determine the mapping $\mathcal{M}_f$ between $\mathcal{S}(q)$ and $\mathcal{F}(q)$ through theoretical modeling and empirical experiments as
\begin{equation}
\mathcal{S}(q) \underset{\text{Modeling}}{\overset{\mathcal{M}_{f}(q)}{\longrightarrow}} \mathcal{F}(q) \underset{\text{Solving}}{\overset{\mathcal{M}_i(q)}{\longrightarrow}} \mathcal{S}(q).
\end{equation}
The essence of single-task sensing is to explore $\mathcal{M}_{f}$ and design the corresponding solving model $\mathcal{M}_i$.

When we extend from single-task sensing to multi-task sensing, the fundamental change lies in the variation of task $q$, which means we need to address the problem of reasoning about the task set $\mathcal{Q}$. Here, $q\in \mathcal{Q}$ represents an individual sensing task.
\begin{equation}\label{eq:multitask}
\mathcal{S}(\mathcal{Q}) \underset{\text{Modeling}}{\overset{\mathcal{M}_{f}(\mathcal{Q})}{\longrightarrow}} \mathcal{F}(\mathcal{Q}) \underset{\text{Solving}}{\overset{\mathcal{M}_i(\mathcal{Q})}{\longrightarrow}} \mathcal{S}(\mathcal{Q}),
\end{equation}
where $\mathcal{S}(\mathcal{Q}) = \bigcup_{i=1}^{n} \mathcal{S}(q_i)$, $\mathcal{F}(\mathcal{Q}) = \bigcup_{i=1}^{n} \mathcal{F}(q_i)$, $\mathcal{M}_{f}(\mathcal{Q}) = \bigcup_{i=1}^{n} \mathcal{M}_{f}(q_i)$ and $q_i \in \mathcal{Q}$.

This equation indicates that multi-task sensing extends single-task sensing by expanding both the event set and feature set.
The new feature set and task set are interconnected through a series of forward models $\mathcal{M}_{f}(q)$. 
The most critical aspect is that the inverse model $\mathcal{M}_i(\mathcal{Q})$ for multi-task sensing is not simply a union of single-task solution models.
When we modify the event set, feature set, or forward model network, we need to redesign the inverse model accordingly, which represents the fundamental bottleneck in multi-task sensing.

In this paper, our tasks, event sets, and feature sets are defined as follows:
1) Tasks: The tasks include tracking, presence detection, and state recognition;
2) Event Set: We define a shared event space $\mathcal{S}$ consisting of low-level human states: absence, stillness (i.e, breathing), local motion, walking, and user locations;
3) Feature Set: We construct a feature set $\mathcal{F}$ by extracting interpretable signal descriptors from raw CSI data.
We focus on three representative features: subcarrier correlation, which captures frequency-domain coherence; DSER, which reflects motion-induced power shifts; and PLCR (path-length change rate), which encodes Doppler-based path variation.
Each feature is computed over multiple temporal windows to capture dynamics at different scales.

For each event in $\mathcal{S}$, we empirically analyze how each feature responds using real-world signal traces. TABLE~\ref{tab:feature-ranges-transpose} summarizes typical ranges. Subcarrier correlation increases with movement intensity and is effective for presence detection; PLCR is sensitive to directional changes and works well for identifying local motion or walking; DSER increases with motion but varies with different window sizes. These features reflect different physical properties and are best used in combination rather than isolation.

\begin{figure}[t]
    \centering
    \includegraphics[width=.9\linewidth]{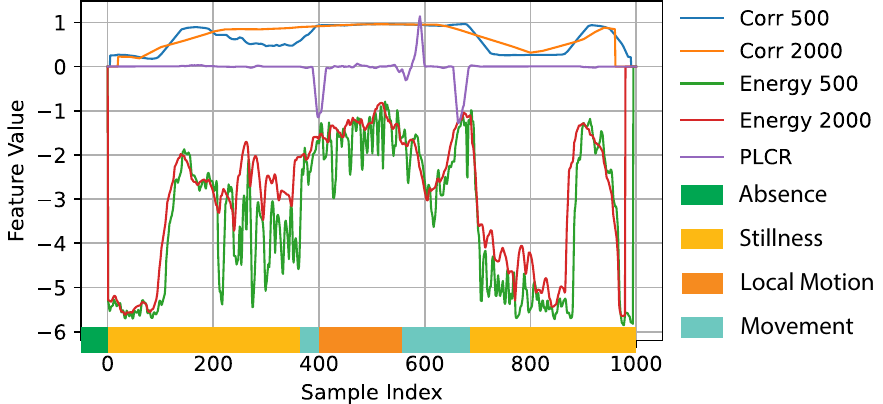}\vspace{-3mm}
    \caption{Comparison between traditional single-task processing and our unified event-flow simulation.}\vspace{-5mm}
    \label{fig:pipeline-compare}
\end{figure}





\begin{algorithm}[t]
\caption{Modeling Track}
\label{algorithm}
\KwIn{Sequence length $L$, event transition matrix $\mathbf M$, boundary $B$, door point $\mathbf d$, sampling rate $f_s$}
\KwOut{Simulated event sequence $\mathcal S$, real event sequence $\mathcal R$, trajectory $\mathcal T$, feature sequence $\mathbf F$}

\BlankLine
\textbf{Step 1: Generate Simulated States and Trajectory}\;
Sample an event pattern from the transition matrix $\mathbf M$\;
Generate simulated event sequence $\mathcal S = [s_1, \ldots, s_L]$\;

\ForEach{contiguous segment $(s,\ell)$ in $\mathcal S$}{
  \uIf{$s$ is a motion state (e.g., walk or rebound)}{
    Assign a direction $\theta$ (randomly or toward $\mathbf d$)\;
    Generate a trapezoidal speed profile with acceleration $a$ and peak speed $v_{\max}$\;
  }
  Propagate the position for $\ell$ time steps with $\Delta t = 1/f_s$ while enforcing boundary $B$\;
}

\BlankLine
\textbf{Step 2: Generate Features}\;
\ForEach{time step $t$ along trajectory $\mathcal T$}{
  Compute Tx--Rx path lengths and the total delay\;
  Compute the range rate and add noise\;
  Extract the feature vector $\mathbf f_t$ (e.g., PLCR)\;
}

\BlankLine
\textbf{Step 3: Map to Real States}\;
\ForEach{time stamp $t \leftarrow 1$ \KwTo $L$}{
  Map $(s_t, v_t, \theta_t)$ to the real state $r_t$ using rule-based mapping\;
  \vspace{-3mm}
}

\BlankLine
\Return{$\mathcal S, \mathcal R, \mathcal T, \mathbf F$}\;
\end{algorithm}

\subsection{Modeling Track: Generate Aggregated Inverse Model}\label{sec:simulation-track}

The primary bottleneck in multi-task aggregation lies in generating the aggregated inverse model.
Our inspiration stems from Eq.~\ref{eq:multitask}.
Specifically, since the event set, feature set, and forward model in multi-task perception all maintain set formulations similar to single-task sensing, we investigate whether we can leverage these known components to design a pipeline for automatically deriving the inverse model.


As shown in Algorithm~\ref{algorithm}, we begin by constructing a sequence of intermediate simulated events that represent human behavioral primitives.
These include (1) leaving through the door, (2) entering the room, (3) walking within the room, (4) remaining still, and (5) performing local motion.
Transitions between these events are governed by a first-order Markov chain, where transition probabilities are chosen to reflect realistic indoor behavior (\textit{e.g.}, short bursts of walking followed by longer periods of stillness).
This abstraction simplifies how presence and absence are modeled over time.

Each simulated event is then expanded into a kinematic profile by sampling human motion parameters.
For walking-related states, we generate velocity and acceleration trajectories with smooth transitions (\textit{i.e.}, starting with acceleration, reaching a peak speed, and decelerating toward the end).
At each step, the direction of motion may change based on a sampled turning probability, allowing the resulting path to approximate natural human trajectories.
For events like entering the room, motion is constrained along a hallway or entry direction.
This process yields a full trajectory aligned with motion characteristics.
Next, we map the simulated event sequence to our defined set of real events (\emph{absence}, \emph{stillness}, \emph{local motion}, \emph{walking}) based on trajectory properties and physical parameters such as speed thresholds and movement extent.
Once labeled, we use the per-event feature models (from the research track) to generate the corresponding feature sequence frame-by-frame.
Each real event emits feature values according to its characteristic distribution, modulated by the motion dynamics in that segment.
The result is a synthetic dataset of variable-length event–feature sequences, denoted $\tilde{\mathcal{D}} = \{(\mathcal{S}_{1:T}, \mathcal{F}_{1:T})\}$. We train a model $\Phi_{\theta}$ to reconstruct the event sequence given its associated feature sequence, effectively learning a task-level solver grounded in interpretable event logic. Compared to direct end-to-end modeling, this design offers better interpretability, label control, and generalization to new task compositions.

\begin{figure*}[!t]
  \subfloat[Scenario 1 (LoS).]{%
    \includegraphics[height=2.9cm]{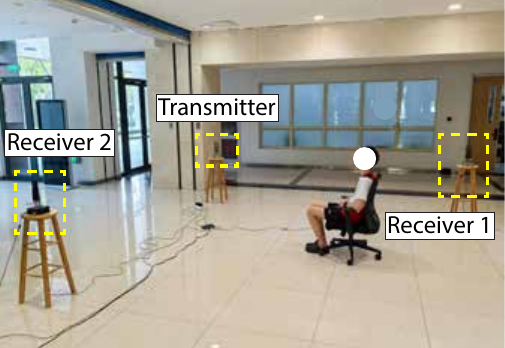}
    \hspace{3pt}
    \includegraphics[height=3.1cm]{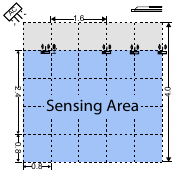}
    \label{fig:env1}
  }
  \hfill
  \subfloat[Scenario 2 (NLoS).]{%
    \includegraphics[height=2.9cm]{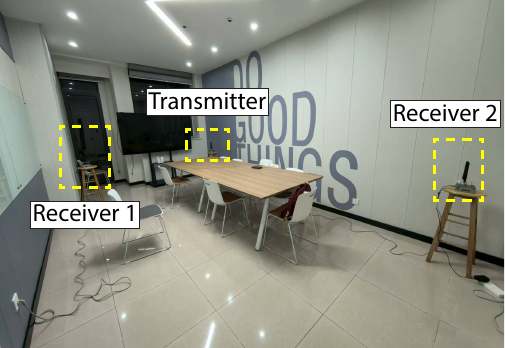}
    \hspace{3pt}
    \includegraphics[height=3.1cm]{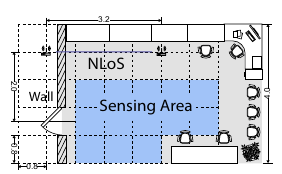}
    \label{fig:env2}
  }
  \caption{Two representative real deployment environments used in our evaluation.}
  \vspace{-.2in}
  \label{fig:implement}
\end{figure*}


\begin{figure*}[t]
\centering
\subfloat[CDF graph of scenario 1.]{\includegraphics[width=0.25\linewidth]{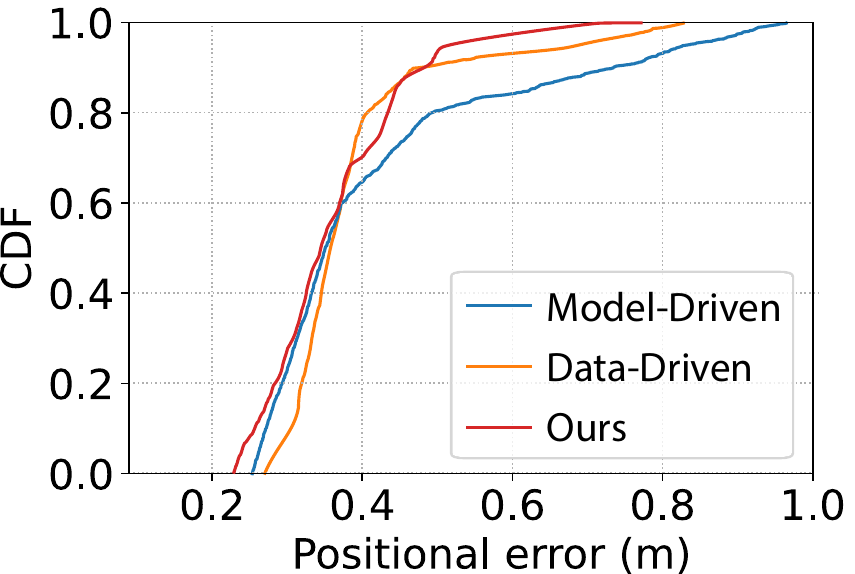}\label{fig:cdf1}}
\subfloat[Confusion matrix of scenario 1.]{\includegraphics[width=0.25\linewidth]{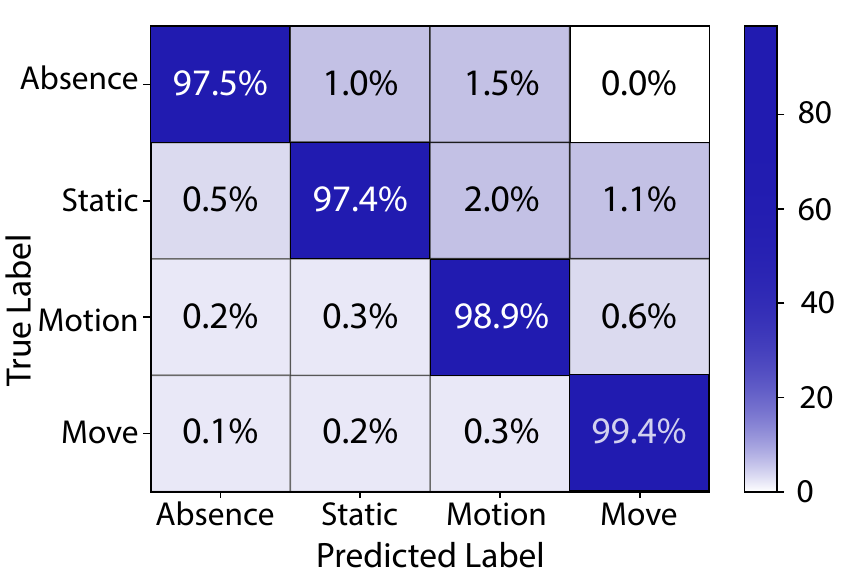}\label{fig:confuse1}}
\subfloat[CDF graph of scenario 2.]{\includegraphics[width=0.25\linewidth]{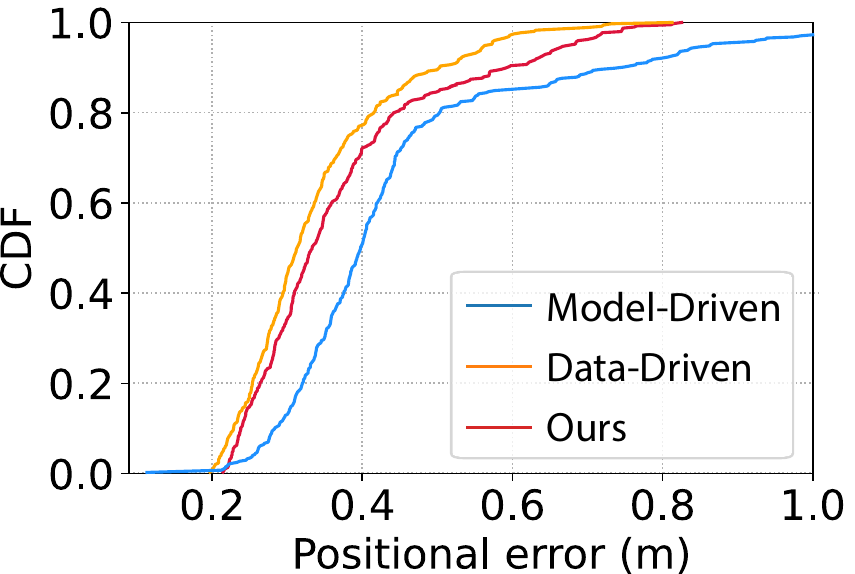}\label{fig:cdf2}}
\subfloat[Confusion matrix of scenario 2.]{\includegraphics[width=0.25\linewidth]{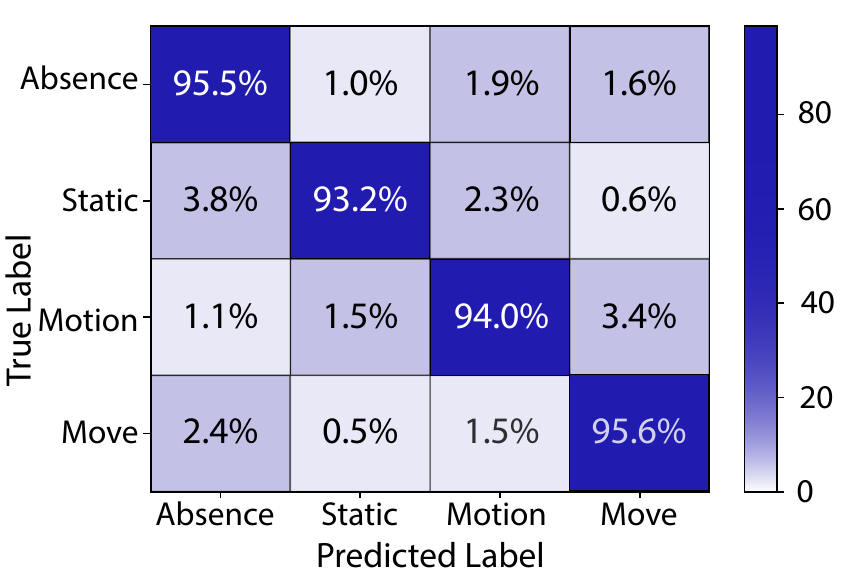}\label{fig:confuse2}}
\caption{Experiments for multi-task performance.}\vspace{-.2in}
\end{figure*}

\subsection{Inference Track: Unified Sensing and Reasoning}\label{sec:inference-path}
The inference track completes the system pipeline by learning how to map observed feature sequences to multiple sensing tasks. Unlike conventional designs that require a dedicated model for each task, our approach enables unified reasoning across tasks using a shared decoder. The goal is to support flexible sensing from any available subset of features.

Given a stream of raw CSI, we first apply the feature extractor (established in the research track) to generate a structured feature sequence. By capturing how key descriptors such as subcarrier correlation and dynamic energy evolve over time, the sequence provides the foundation for inference. The extracted features are then fed into a task-level model learned in the modeling track, which has been trained to associate feature dynamics with event transitions.

The unified latent representation is passed to a temporal decoder, implemented as a lightweight Transformer, which models temporal context. A multi-head classifier then produces per-frame event probabilities for all supported tasks, such as detecting breathing, walking, or transitions between states. Since all predictions are made in a shared space, the decoder supports multi-task inference without duplicating model components.
\section{Evaluation}\label{sec:experiment}
\subsection{Experimental Setup}\label{sec:exp-setup}

\begin{figure*}
\subfloat[Tracking results of Uni-Fi.]{%
  \includegraphics[width=0.16\linewidth]{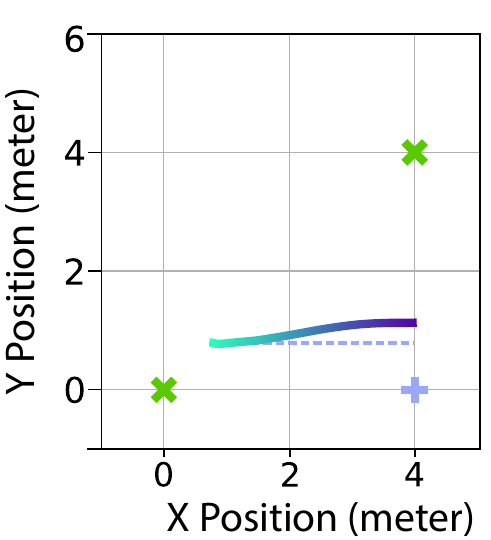}
  \includegraphics[width=0.16\linewidth]{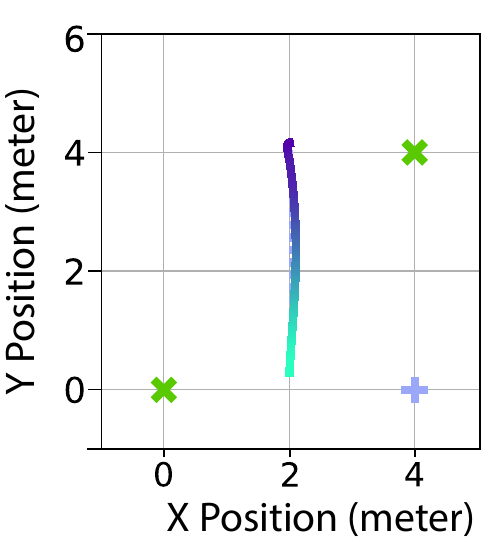}
  \includegraphics[width=0.16\linewidth]{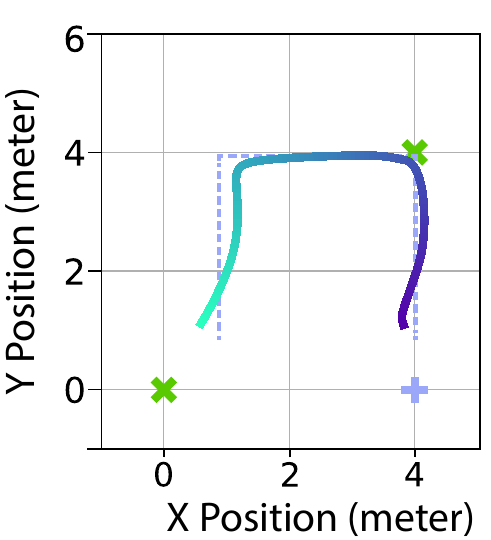}
  \includegraphics[width=0.16\linewidth]{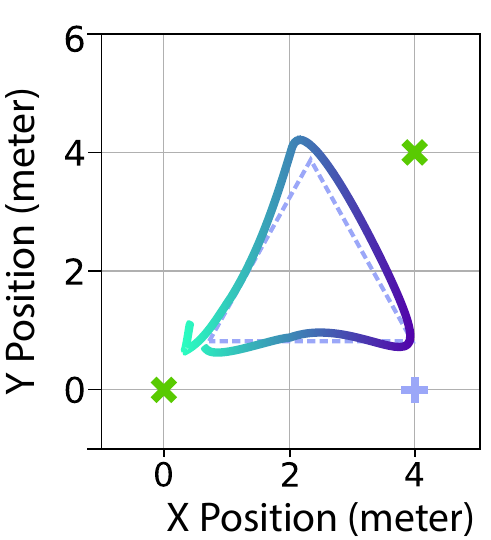}
  \includegraphics[width=0.16\linewidth]{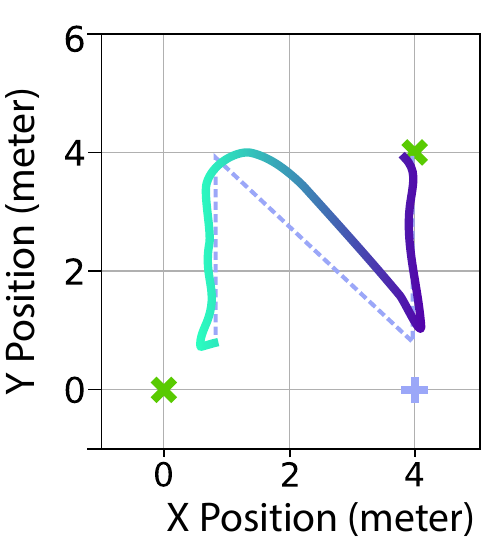}
  \includegraphics[width=0.16\linewidth]{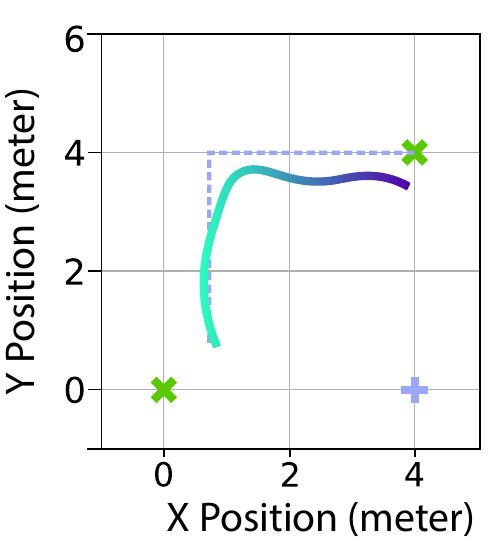}
  \label{fig:unifi}}
  \vspace{-2mm}
\centering
\subfloat[Tracking results of data-driven approach.]{%
  \includegraphics[width=0.16\linewidth]{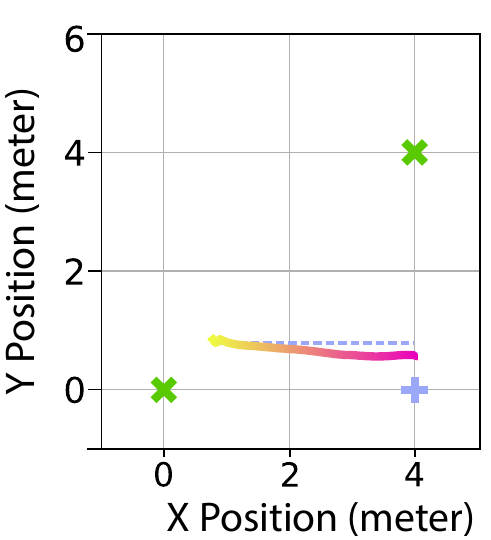}
  \includegraphics[width=0.16\linewidth]{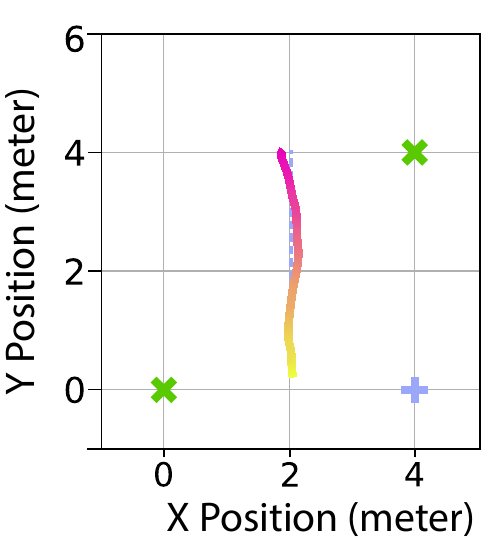}
  \includegraphics[width=0.16\linewidth]{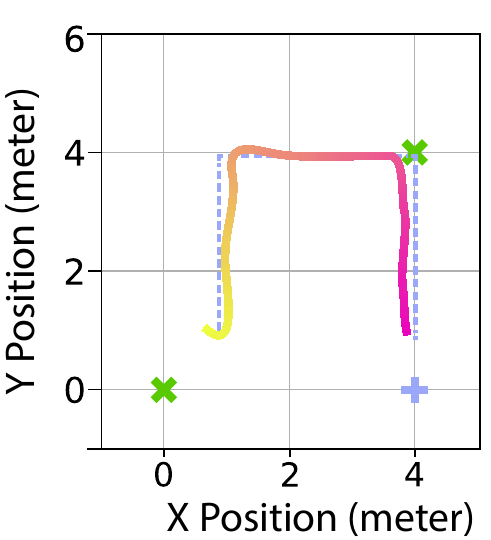}
  \includegraphics[width=0.16\linewidth]{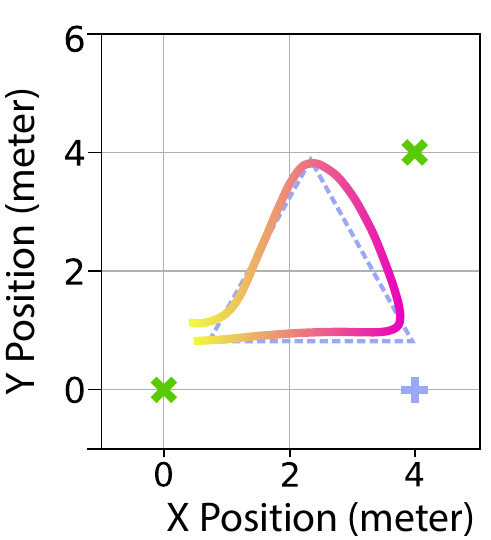}
  \includegraphics[width=0.16\linewidth]{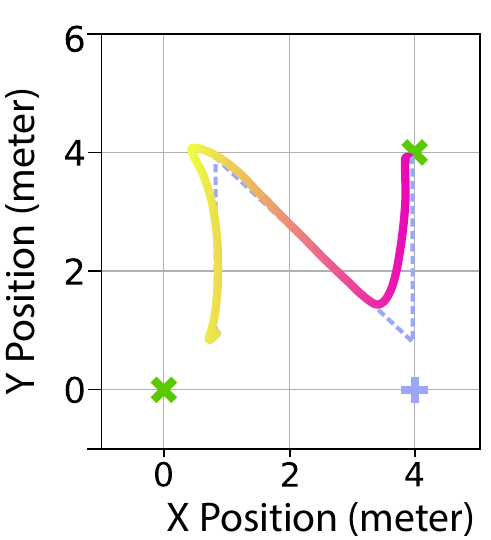}
  \includegraphics[width=0.16\linewidth]{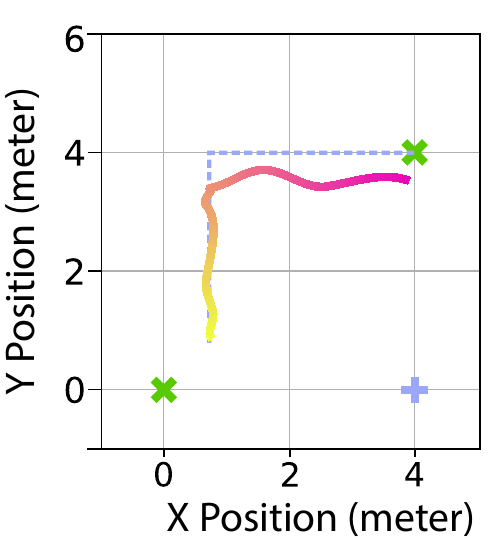}
\label{fig:nne}}
  \vspace{-2mm}
\centering
\subfloat[Tracking results of model-driven approach.]{%
  \includegraphics[width=0.16\linewidth]{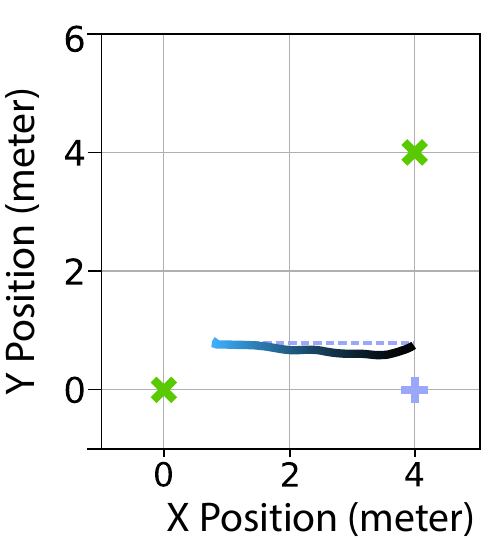}
  \includegraphics[width=0.16\linewidth]{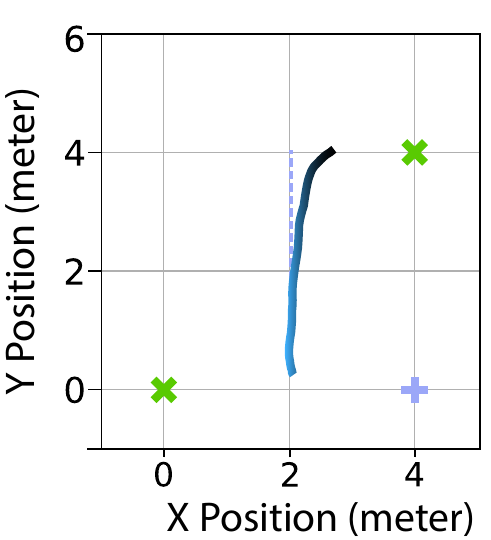}
  \includegraphics[width=0.16\linewidth]{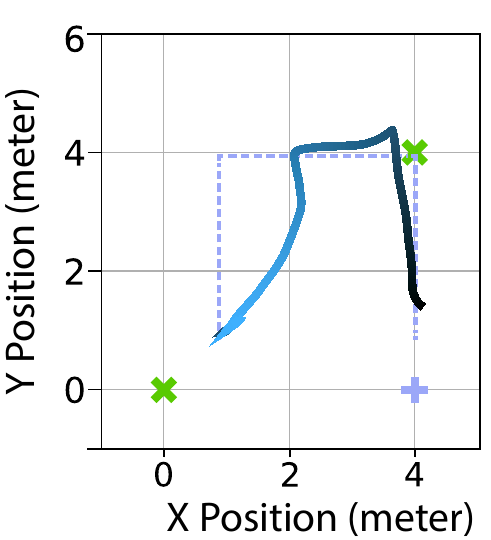}
  \includegraphics[width=0.16\linewidth]{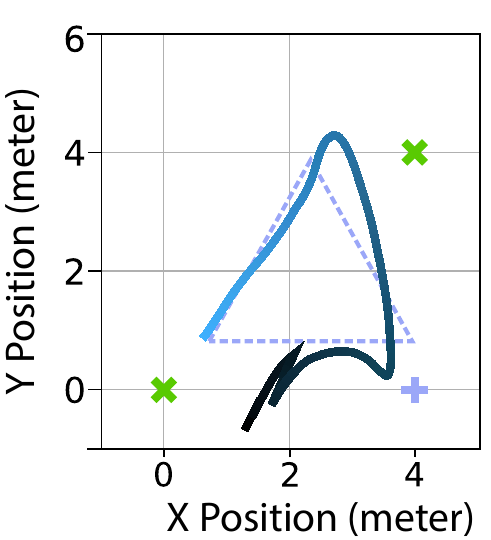}
  \includegraphics[width=0.16\linewidth]{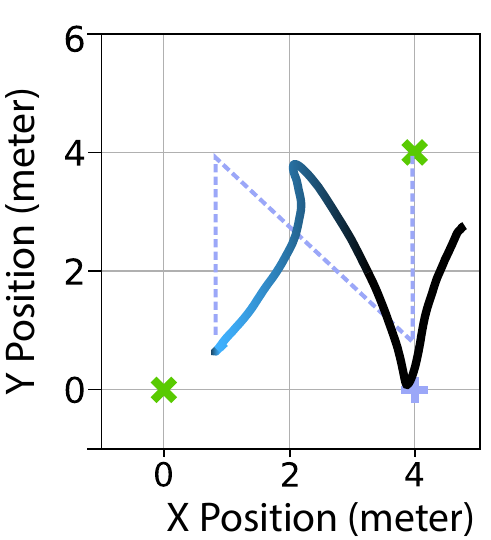}
  \includegraphics[width=0.16\linewidth]{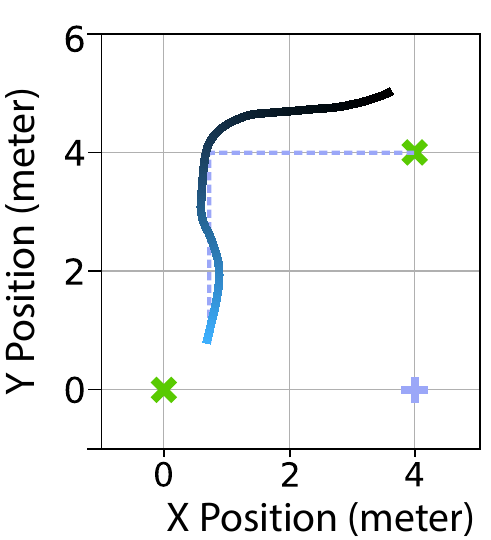}
  \label{fig:widar}
  \vspace{-4mm}
}
\caption{Comparison of the tracking performance among \textsc{Uni-Fi}, \textit{NNE-Tracking}, and \textit{Witraj}.}\vspace{-5mm}
\label{difobs-track}
\end{figure*}

\subsubsection{Machine Configuration}
We conduct experiments using three laptops, each running Ubuntu 16.04 and equipped with Intel 5300 network interface cards (NICs). The wireless setup consists of one transmitter and two receivers. The transmitter uses a single directional antenna, while each receiver is equipped with a linear array of three antennas spaced $2.5\,\mathrm{cm}$ apart. All devices operate in the $5.31$–$5.33\,\mathrm{GHz}$ band.

\subsubsection{Scene Arrangement}
To evaluate performance under realistic conditions, we test \textsc{Uni-Fi} in two indoor environments with differing layouts and scales. As shown in Fig.~\ref{fig:env1}, the first setting is an open space, whereas Fig.~\ref{fig:env2} shows a meeting room with more multipaths. Evaluating the system in these distinct settings demonstrates its reliable performance in both compact environments and open areas. Both spaces emulate typical office or home environments. These settings introduce varying degrees of occlusion, multipath, and layout complexity, allowing us to assess the robustness and adaptability of \textsc{Uni-Fi} in dynamic indoor scenarios.

\subsection{Multi-task Performance}\label{sec:exp-multitask}
\subsubsection{Event Classification}
We evaluate the performance of \textsc{Uni-Fi} on the task of frame-wise event classification, distinguishing between absence, stillness, local motion, and walking. The system processes raw CSI streams, extracts a multi-window feature sequence, and applies the shared inference module trained with both real and simulated sequences.

Each recording session includes annotated transitions between different behavioral states. No environment-specific tuning is applied during training or inference.
To assess performance, we compute per-frame classification accuracy and confusion matrices under different datasets. Results are shown in Fig.~\ref{fig:confuse1} and Fig.~\ref{fig:confuse2}. \textsc{Uni-Fi} consistently achieves high accuracy across both settings, demonstrating robustness to environment variation.

\subsubsection{Device-free Tracking}

We evaluate \textsc{Uni-Fi}'s ability to estimate human trajectories in a fully device-free setting, without using any ground-truth coordinate annotations during training. In contrast to existing systems that rely on supervised trajectory labels, our approach learns from structured feature dynamics derived from event-aware simulation and real-world signal patterns.

Fig.~\ref{difobs-track} visualizes the tracking results of three methods on six identical trajectories: a classical model-driven baseline~\cite{qian2017widar}, a recent data-driven approach~\cite{tong2024nne}, and our proposed system. As shown in Fig.~\ref{fig:unifi}, \textsc{Uni-Fi} produces smooth and accurate paths that closely follow the true movement, despite the absence of explicit coordinate supervision. The data-driven baseline (Fig.~\ref{fig:nne}) achieves comparable accuracy but requires extensive task-specific labeling. In contrast, the model-driven method (Fig.~\ref{fig:widar}) suffers from trajectory drift and poor resolution, particularly in cluttered environments.
This result is not achieved through end-to-end trajectory fitting. Instead, \textsc{Uni-Fi} relies on a compact set of physical features such as subcarrier correlation and DSER to infer the underlying structure of human motion. These features help detect transitions, pauses, and changes in movement direction, and remain reliable even in environments with strong multipath effects.

We further evaluate statistical performance using cumulative distribution functions (CDFs) in two distinct environments. As shown in Fig.~\ref{fig:cdf1} and Fig.~\ref{fig:cdf2}, \textsc{Uni-Fi} consistently outperforms the baselines across different spatial layouts. At the 90th percentile, the model-driven approach yields a localization error of up to 0.72\,m, while the data-driven method reduces this to 0.50\,m. \textsc{Uni-Fi} achieves the best performance, with 90\% of errors falling below 0.43\,m.

These results demonstrate the advantage of modeling feature-level dynamics rather than directly regressing coordinates, especially in label-scarce, device-free scenarios.

\begin{figure*}[t]
\hfill
\subfloat[Training epochs.]{\includegraphics[width=0.25\linewidth]{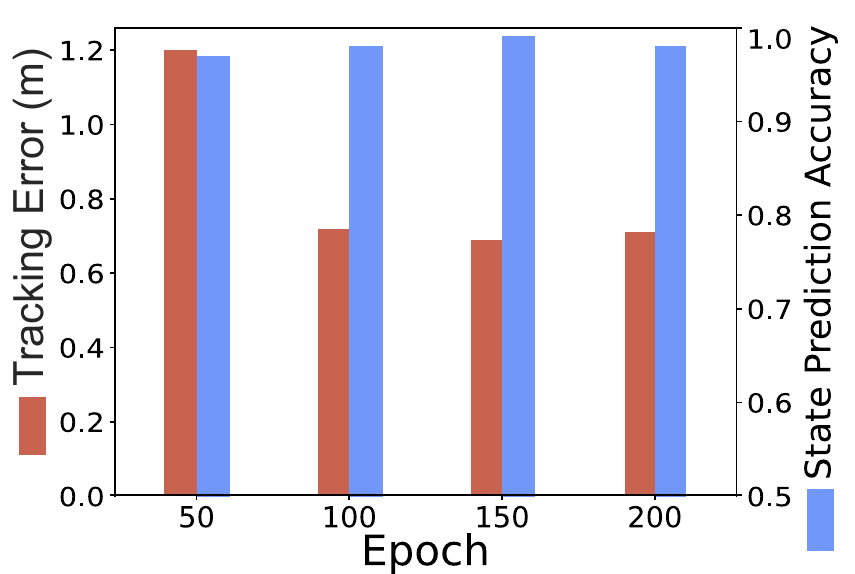}\label{fig:epoch}}\hfill
\subfloat[Neural network structures.]{\includegraphics[width=0.25\linewidth]{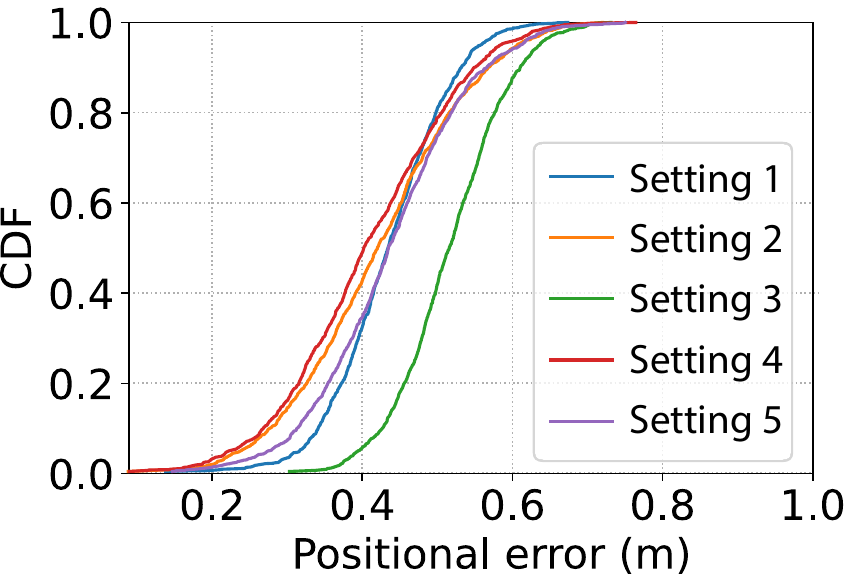}\label{fig:cdf}}\hfill
\subfloat[Packet Rates.]{\includegraphics[width=0.25\linewidth]{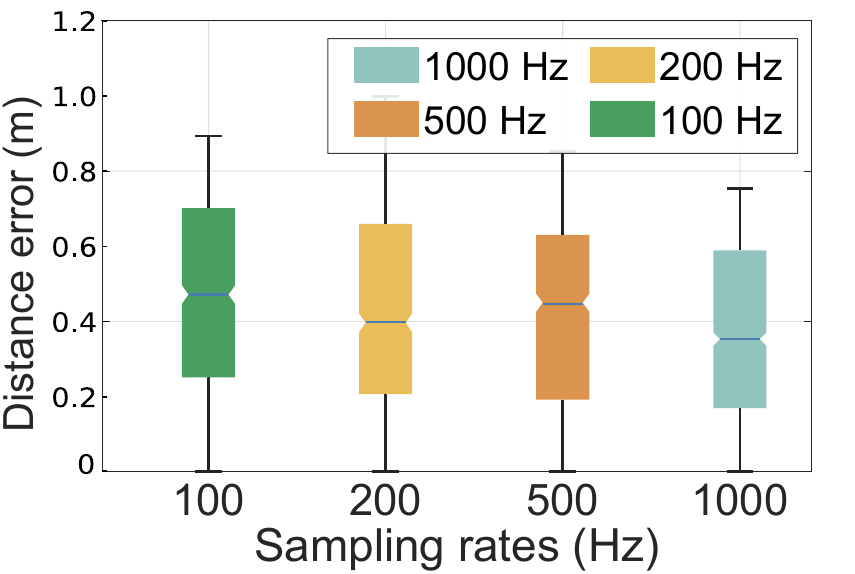}\label{fig:packetrate}}\hfill
\subfloat[Runtime.]{\includegraphics[width=0.25\linewidth]{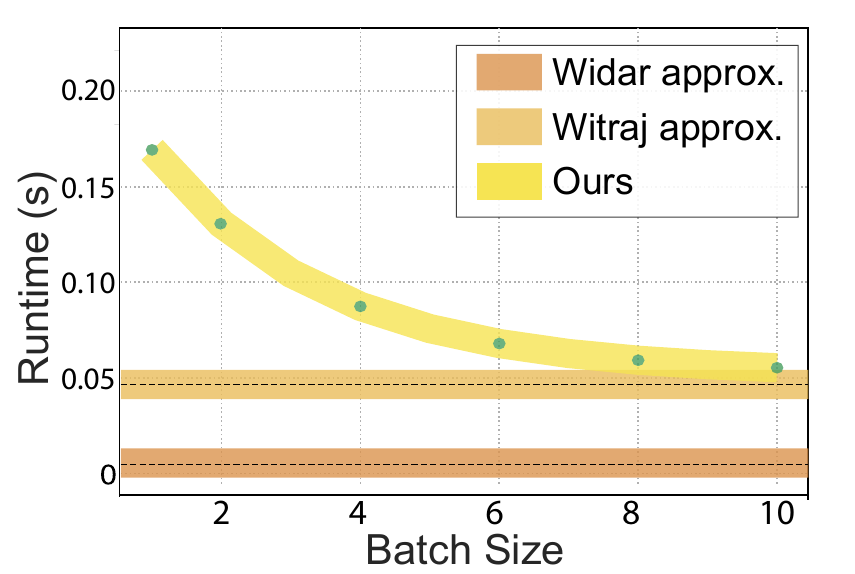}\label{fig:runtime}}
\caption{Experiments for generalization analysis.}\vspace{-3mm}
\end{figure*}



\subsection{Generalization Analysis}\label{sec:exp-generalization}

To evaluate the generalization capability of \textsc{Uni-Fi}, we conduct a series of controlled experiments by varying key modeling factors: training epochs, network architecture, packet rates, and system runtime.

\textbf{Training epochs.}  
We vary the number of training epochs from 50 to 200 to examine stability. As shown in Fig.~\ref{fig:epoch}, most models stabilize after 100 epochs, with diminishing returns beyond that. Notably, the accuracy gap between early and late training becomes narrower when simulation-augmented data is used. This suggests that synthetic supervision improves training efficiency and accelerates convergence.

\textbf{Model architectures.}
We evaluate five configurations by varying the hidden dimensions and training settings. The state encoder uses 128, 256, or 512 units, and the trajectory encoder uses 128 or 256. All models are trained with batch size 128, learning rate $2\times10^{-3}$, and equal loss weights ($\lambda_{\text{pos}} = \lambda_{\text{sta}} = 1.0$), unless otherwise specified.
Setting 1 (256/128) is the default. Settings 2 and 3 change the state encoder to 128 and 512. Setting 4 increases the trajectory encoder to 256. Setting 5 lowers the learning rate to $1\times10^{-3}$ and sets $\lambda_{\text{pos}}=0.5$, $\lambda_{\text{sta}}=1.5$ to test training sensitivity.
As shown in Fig.~\ref{fig:cdf}, Setting 4 achieves the best median classification accuracy. The LSTM baseline shows larger variance and weaker generalization. In tracking, the best Transformer yields a median localization error of $0.52\,\mathrm{m}$, versus $0.68\,\mathrm{m}$ for the LSTM-based model.

\textbf{Packet Rates.}  
The packet rate determines the temporal resolution of CSI sampling and affects the system’s ability to capture motion dynamics. Lower rates may miss brief transitions or subtle movements. We evaluate \textsc{Uni-Fi} under four rates: 100, 200, 500, and 1000\,Hz.
As shown in Fig.~\ref{fig:packetrate}, the system achieves average tracking errors of $0.58\,\mathrm{m}$ at $100\,\mathrm{Hz}$ and $0.52\,\mathrm{m}$ at $1000\,\mathrm{Hz}$, with $0.56\,\mathrm{m}$ and $0.54\,\mathrm{m}$ at $200$ and $500$\,Hz, respectively. While higher sampling rates improve resolution, the accuracy gain plateaus beyond $500$\,Hz. This indicates that \textsc{Uni-Fi} effectively captures motion with stable, low-rate features, enabling reliable tracking even on hardware with constrained sampling capacity.

\textbf{Runtime analysis.}  
We evaluate the real-time feasibility of \textsc{Uni-Fi} by measuring the average inference latency per sample on a standard CPU. As shown in Fig.~\ref{fig:runtime}, it achieves an average processing time of 0.05\,s per sample, which is well within the requirements for real-time human behavior sensing. This confirms that \textsc{Uni-Fi} can be deployed in practical scenarios without the need for specialized hardware acceleration.

\section{Related Work}\label{sec:related}
In this section, we organize recent work into three typical sensing tasks including presence detection, activity classification, and user tracking.
\subsection{Presence Detection}
Presence detection has emerged as a fundamental capability for smart environment control and intrusion detection systems.
The core principle involves identifying human presence through respiration patterns or body movements that modulate Wi-Fi signals.
Recent advances in wireless sensing have enabled various device-free presence detection systems using commercial Wi-Fi signals.
Jayaweera et al.~\cite{jayaweera2024robust} propose a robust in-car Child Presence Detection (CPD) system; their system leverages breathing spectrum enhancement and CNN-based classification to minimize false alarms.
For multi-room scenarios, TCD-FERN~\cite{shen2024time} develops a time-selective conditional dual-feature recurrent network achieving 97.35\% accuracy through dynamic-static data preprocessing and voting schemes.
Their system uniquely addresses both line-of-sight and non-line-of-sight conditions in vehicular environments.
\subsection{Activity Classification}
Recent  Wi-Fi-based human activity recognition (HAR) have demonstrated the effectiveness of modeling and machine learning techniques for extracting meaningful patterns from CSI. The core challenge across these approaches lies in extracting suitable features from the raw CSI measurements. 
The model-based approaches demonstrate how carefully designed physical models can extract robust motion features from noisy CSI measurements. For instance,
Widar3.0~\cite{zhang2021widar3} introduces body-coordinate velocity profiles (BVPs) as domain-independent features and uses matching networks to learn transferable representations across environments.
The data-driven approaches eliminate the need for manual feature engineering by automatically learning discriminative patterns from CSI sequences.
WiLife~\cite{li2025wilife} partitions living spaces into functional areas and categorizes daily activities into atomic states (still, in-situ moving, walking) by analyzing Doppler effects in CSI signals.

\subsection{User Tracking}
User tracking has become increasingly important, with user location serving as a key contextual input for enabling such services. The core principle involves extracting velocity features from Wi-Fi signals that reflect human movement patterns for continuous tracking.
NNE-Tracking\cite{tong2024nne} proposes a paradigm to fuse data-and model-based methods. It leverages a neural network as a fast data-based solver and integrates a model-based supervisor to evaluate trace predictions.
WSTrack~\cite{tian2023wstrack} first introduces a Wi-Fi-acoustic fusion model that leverages the interaction between Fresnel zones and footstep sound rays to track a silent user.
MetaTrack~\cite{meng2025metatrack} defines a method for constructing Wi-Fi sensing zones in complex scenarios, laying a theoretical foundation for user tracking in such environments.
For multi-user scenarios, WiPolar~\cite{venkatnarayan2020leveraging} enables simultaneous tracking of multiple people by leveraging the polarization characteristics of Wi-Fi signals. 
WiMap~\cite{tan2025wimap} can construct a device topology map without requiring calibration, providing a solid foundation for plug-and-play user tracking.

To summarize, recent Wi-Fi sensing research has steadily improved robustness and generalization across environments by combining physics-guided representations with data-driven learning and by expanding from single-task settings to multi-room and multi-user deployments.

\section{Conclusion}\label{sec:conclude}

In this paper, we presented \textsc{Uni-Fi}, a unified Wi-Fi sensing framework designed for sustainable and multi-task deployment. By combining temporal statistics with spectral--spatial coherence, \textsc{Uni-Fi} provides a robust feature representation than raw CSI streams. Its architecture further supports seamless feature evolution and task adaptation without requiring full-model retraining.
Extensive experiments across representative sensing tasks demonstrate the framework's superior performance. Specifically, \textsc{Uni-Fi} achieves a high-precision localization error of approximately 0.54m, while maintaining exceptional classification accuracies of 98.34\% for activity recognition and 98.57\% for presence detection. These empirical results confirm that multi-feature integration consistently and substantially improves sensing stability and scalability under diverse conditions. Overall, \textsc{Uni-Fi} establishes a robust and unified framework for practical multi-task Wi-Fi sensing, demonstrating the efficacy of feature-centric integration for scalable sensing systems.




\twocolumn[
  \begin{@twocolumnfalse}
    \centering
    \section*{References}
    \vspace{1em} 
  \end{@twocolumnfalse}
]
\balance
\begingroup
\renewcommand{\section}[2]{}%
\bibliography{references}
\endgroup
\bibliographystyle{ieeetr}
\end{document}